\documentclass[%
 aip,
 floatfix,
 amsmath,amssymb,
 reprint,%
]{revtex4-1}

\usepackage{graphicx}
\usepackage{amsfonts, amsmath, amstext, amssymb, amsfonts, amsxtra}
\usepackage{braket}
\usepackage{bbm}
\usepackage{bbold}
\usepackage{dcolumn}
\usepackage[protrusion=true,expansion=true]{microtype}
\usepackage{wasysym}
\usepackage{cleveref}
\usepackage{color}
\usepackage{ulem}
\usepackage{bm}
\usepackage[utf8]{inputenc}
\usepackage[T1]{fontenc}
\usepackage{mathptmx}
\usepackage{float}

\newcommand{\im}{{i}}         % imaginary unit
\begin{document}

\title{Chaotic spin-photonic quantum states \\in an open periodically modulated cavity}
% Force line breaks with \\

\author{I.I.~Yusipov$^1$}
\email{yusipov.igor@gmail.com}
\author{S.V.~Denisov$^{2}$, and M.V.~Ivanchenko $^1$}
\affiliation
{ 
    $^1$ Department of Applied Mathematics, Lobachevsky University, Nizhny Novgorod, Russia \\
    $^2$ Department of Computer Science, Oslo Metropolitan University, Oslo, Norway
}

\date{\today}% It is always \today, today,

\begin{abstract}
When applied to dynamical systems, both classical and quantum, time periodic modulations can produce complex non-equilibrium states which  are often termed 'chaotic`.
Being well understood within the unitary Hamiltonian framework, this phenomenon is  less  explored in open quantum systems. 
Here we consider quantum chaotic state emerging in a leaky cavity, when an intracavity photonic mode is coherently pumped with the intensity  varying  periodically in time. 
We show that a single spin, when placed inside the cavity and coupled to the mode, can moderate  transitions between  regular and chaotic regimes -- that are identified by using quantum Lyapunov exponents -- and thus can be  used to control the degree of chaos.   
In an experiment, these transitions can be detected by analyzing photon emission statistics. 
\end{abstract}

\maketitle

\begin{quotation}
A passage connecting Chaos Theory \cite{ott_2002} and many-body quantum physics is provided by the mean-field ideology \cite{spohn_1980, kadanoff_2007, breuer_petruccione_2010} and different semiclassical approximations \cite{altland_haake_2012, chavez-carlos_bastarrachea-magnani_2016}. 
They declare that, when the number $N$ of quantum degrees of freedoms is systematically increased, the exponentially complex evolution of a quantum model can be approximated  with a fixed size system of classical non-linear differential equations.
These equations model the dynamics of the expectation values of relevant observables and the model becomes  exact in the thermodynamic limit $N \rightarrow \infty$.
A degree of chaos in the original quantum system can be quantified by calculating  standard classical quantifiers (usually maximal Lyapunov exponents \cite{chavez-carlos_bastarrachea-magnani_2016, chavez-carlos_lopez-del-carpio_2019})  for the corresponding classical system.
In the case of an open quantum system, the mean-field approach can be realized on the level of density matrices \cite{schiro_joshi_bordyuh_2016}.
Alternatively, the adjoint form of a Markovian master equation, governing the evolution of the system density matrix \cite{breuer_petruccione_2010}, can be employed \cite{xu_tieri_2014, hartmann_poletti_2017, patra_altshuler_2019}.
% The adjoint equation determines the evolution of observables and allows one to obtain a hierarchy of coupled classical equations describing dynamics of the  expectation %values of relevant observables and of their correlators.
What if we are dealing with an open model and do not want (or simply do not have a possibility) 
to go into the (semi)classical limit or resort to a mean-field description? 
When the evolution of the system is modeled with a master equations of the Gorini-Kossakowski-Sudarshan-Lindblad (GKSL) form \cite{breuer_petruccione_2010},  a recently proposed idea of quantum Lyapunov exponents \cite{yusipov_vershinina_2019} provides a possibility to quantify 
the degree of chaos in a straightforward manner.
Here we implement this idea and demonstrate how chaotic regimes of a system with $N \gg 1$ states (a photonic mode in an open cavity) can  be controlled by coupling it to a single spin.
\end{quotation}

\section{Introduction}
% In real-life realizations of  nano- and opto-mechanical systems \cite{Poot2012,Aspelmeyer2014} and superconductive circuits \cite{Clarke2008} their complete isolation from surrounding environment is practically impossible. 
% The ideal coherent quantum evolution is only an approximation of the experimental reality, whose validity is restricted to certain timescales.
% The asymptotic states of these systems are sculpted not only by their internal dynamics but also by the action of their environments. 
% The resulting effects, commonly referred to as ``dissipation'', drive a quantum system to an asymptotic state, a quantum attractor, that can be no less complex that its Hamiltonian predecessor \cite{Breuer2002,Zoller2008,Zoller2011}. 
% The other degree of control can be obtained by time-periodic modulations that in case of coherent systems can create a set of non-equilibrium eigenstates. 
% Known as Floquet states, they are distinctly different from the quantum states exhibited by the same system in the stationary limit \cite{Sambe1973,Hanggi1998}.
% The interplay between decoherence and periodic modulation creates novel quantum attractor states, which understanding is still far from complete \cite{Hartmann2017,Ivanchenko2017,Poletti2017,Poletti2018,Yusipov2019a}.
Solid state cavity QED systems \cite{walther_varcoe_2006, arakawa_finley_2015} are an appealing choice -- both as a theoretical framework and experimental test-beds -- to investigate Quantum Chaos in open systems. 
% Because of their flexibility and ability to reproduce in vitro quantum effects at relatively low cost (as compared to similar quantum optical setups). 
Recent advances in the field of solid-state technologies allow, e.g.,  to fabricate a single semiconductor quantum dot and embed it into a microcavity  \cite{arakawa_finley_2015}. 
The corresponding quantum dot can have from two to four energy levels, with transition frequencies ranging from the infrared to ultraviolet ends of the electromagnetic spectrum. 
Interaction  between cavity modes and dot excitons can also be tuned \cite{reithmaier_sek_2004, hennessy_badolato_2007}. 
Small number of energy levels makes quantum dots good candidates to realize qubits (or qudits), while interaction between different qubits can be mediated by the cavity mode.
A basic cavity QED model typically includes a two-level system, a 'spin',  placed inside a cavity and coupled to the photonic mode.
	
In our recent work \cite{yusipov_vershinina_2020} we showed that a photonic mode of an open and periodically modulated Kerr-nonlinear cavity can exhibit transitions from regular dynamics to chaos. 
A degree of chaos is quantified with quantum Lyapunov exponent \cite{yusipov_vershinina_2019}. 
These transitions are also associated with modification of the probability distribution of photon emission waiting times, which changes its intermediate asymptotic from the exponential (regular dynamics) to a power-law (chaos) decay. 
	
In this paper we consider spin-photonic states, emerging in a single-spin cavity QED model. 
We demonstrate that, by tuning spin-photon coupling, we  can modify the photonic mode dynamics and induce transitions from regular to chaotic regimes.  

\section{Model}
We consider a photonic mode in a leaky Kerr-nonlinear cavity which is periodically pumped by an external coherent electromagnetic field \cite{spiller_ralph_1994, brun_percival_1996}.
The model is described by the Hamiltonian
\begin{equation}
    H(t) = H_{\mathrm{s}} + H_{\mathrm{ph}}(t) + H_{\mathrm{int}},
    \label{eq:Hamiltonian}
\end{equation}
where Hamiltonians
\begin{equation}
    \begin{gathered}
    H_{\mathrm{s}} = \frac{\delta}{2}J_z,\\
    H_{\mathrm{ph}}(t) = \frac{1}{2} \chi  a^\dagger a^\dagger a a + i F(t) \left( a^\dagger - a \right), \\
    H_{\mathrm{int}} = \frac{g}{2}\left( a^\dagger J_{-} + J_{+} a \right).    
    \label{eq:Hamiltonian_Parts}
    \end{gathered}
\end{equation}
describe the dynamics of the spin, of the mode, %$H_{ph}$%, 
and the interaction between them, respectively. %$H_{int}$. 
Here $\chi$ is the photon interaction strength (an effective non-linearity parameter), $\hat{a}^\dagger$ and $\hat{a}$ are photon creation and annihilation operators, and $\hat{n}=\hat{a}^{\dagger}\hat{a}$ is the photon number operator. 
Modulation function $F(t)=F(t+T)$ models a bi-valued quench-like driving of period $T$; more specifically, $F(t) = A$ within $0 < t \le T/2$ and $F(t) = 0$ for the second half period $ T/2 < t \leq T$. 
Finally, $J_{z}$, $J_{+}$, $J_{-}$ are spin operators, $\delta$ is detuning of the resonant frequency of spins from the frequency of the optical mode, and $g$ is the strength of the spin-photon coupling.

The evolution of the total system is modeled with the Lindblad master equation (henceforth we set $\hbar = 1$) \cite{breuer_petruccione_2010, alicki_lendi_2007}:
\begin{equation}
    \begin{aligned}
    \dot{\varrho} = \mathcal{L}(\varrho) = -\im [H(t),\varrho] + \mathcal{D}(\varrho),
    \label{eq:master}
    \end{aligned}
\end{equation}
where the first term on the r.h.s.\ captures the unitary evolution of the system, determined by Hamiltonian (\ref{eq:Hamiltonian}, \ref{eq:Hamiltonian_Parts}), while the second term describes a dissipative interaction with the environment. 

There are two dissipative channels. 
First, photons can be emitted from the cavity and the rate of this process is specified by constant $\gamma$. 
Second, there is spontaneous spin relaxation to the ground state; the rate of this process is determined by the constant $w$. 

Accordingly, dissipation is modeled with two Liouville operators,
\begin{equation}
    \begin{gathered}
    \mathcal{D}(\varrho) = L_{\mathrm{ph}}(\varrho) + L_{\mathrm{s}}(\varrho),\\
    L_{\mathrm{ph}}(\varrho) = \gamma \left( a \varrho a^{\dagger} - \frac{1}{2} a^{\dagger} a \varrho - \frac{1}{2} \varrho a^{\dagger} a \right),\\
    L_{\mathrm{s}}(\varrho) = \omega \left( \sigma^{-} \varrho \sigma^{+} - \frac{1}{2} \sigma^{+} \sigma^{-} \varrho - \frac{1}{2} \varrho \sigma^{+} \sigma^{-} \right).
    \label{eq:dissipation}
    \end{gathered}
\end{equation}

In numerical simulations, we limit the number of photons in the cavity mode by integer $N$ so that the Hilbert space of the total system  has dimension $2 (N+1)$.
Parameter $N$ is chosen to be large enough so that the average number of photons in the cavity, $\langle N_{\mathrm{ph}} \rangle$, is substantially smaller than $N$.
$\langle N_{\mathrm{ph}} \rangle$ depends on parameters of Hamiltonian; yet the main control parameter, which determines the mean number of photons, is coupling strength $\chi$ \cite{spiller_ralph_1994, brun_percival_1996}.
Throughout the paper we set $\chi=0.008$, $\gamma=0.1$. 
It corresponds to $\langle N_{\mathrm{ph}} \rangle \sim 50$ and we set $N=300$.

\section{Methods}
In simulations, we use quantum Monte-Carlo wave function method to unravel  deterministic  equation (\ref{eq:master}) into an ensemble of quantum trajectories \cite{dum_parkins_1992, molmer_castin_1993, plenio_knight_1998, daley_2014}. 
It allows for describing the evolution of the model system in terms of ensemble of pure states, $\psi(t)$, governed by an effective non-Hermitian Hamiltonian \cite{spiller_ralph_1994, brun_percival_1996},
\begin{equation}
    \begin{aligned}
    \dot{\psi}=H(t)\psi -\frac{i}{2} \sum_{k=\mathrm{s,ph}}V_k^\dagger V_k\psi,
    \label{eq:nonHermHamiltonian}
    \end{aligned}
\end{equation} 
where $V_\mathrm{s} = w\sigma^{-}$ and $V_\mathrm{s} = \gamma a$. 	
The norm of the wave function decays according to 
\begin{equation}
    \begin{aligned}
    \frac{d}{dt}||\psi||= -\sum_{k=\mathrm{s,ph}}\psi^* V_k^\dagger V_k\psi,
    \label{eq:normDecay}
    \end{aligned}
\end{equation} 
and as it reaches a threshold $\eta$, repeatedly drawn as i.i.d. random number from the unit interval $[0,1]$, a random jump is performed \cite{molmer_castin_1993}, and the norm is reset to $||\psi(t)||=1$. 
Then a round of the continuous non-unitary evolution,  Eq.~(\ref{eq:normDecay}), is repeated again, until the next quantum jump occurs, etc.  
For the model given by Eqs.(\ref{eq:Hamiltonian},\ref{eq:master},\ref{eq:dissipation}), a quantum jump that corresponds to an emission of a single photon can be detected  with a photodetector \cite{carmichael_1993}. 

The density matrix can then be sampled from a set of $M_r$ realizations as $\varrho(t_\mathrm{p};M_{\mathrm{r}}) = \frac{1}{M_{r}} \sum_{j=1}^{M_{\mathrm{r}}} \ket{\psi_j(t_\mathrm{p})}\bra{\psi_j(t_\mathrm{p})}$, which, given an initial pure state $\psi^\mathrm{init}$ for Eq.~(\ref{eq:nonHermHamiltonian}), converges towards the solution of Eq.~(\ref{eq:master}) at time $t_\mathrm{p}$ for the initial density matrix $\varrho^{\mathrm{init}} = \ket{\psi^\mathrm{init}}\bra{\psi^\mathrm{init}}$. 

Following Refs.~\cite{spiller_ralph_1994, brun_percival_1996}, we make use of the complex-valued observable of the non-Hermitian photon annihilation operator:  
\begin{equation}
    \begin{aligned}
    \theta(t)&=\langle\psi^\dagger(t)| a |\psi(t)\rangle.
    \label{eq:theta}
    \end{aligned}
\end{equation}
Additionally, we calculate the following observables for the spin subsystem:
\begin{equation}
    \begin{gathered}
    \nu(t) = \langle\psi^\dagger(t)| J_+ |\psi(t)\rangle,\\
    \eta(t) = \langle\psi^\dagger(t)| J_z |\psi(t)\rangle.
    \label{eq:eta_nu}
    \end{gathered}
\end{equation}

To calculate the largest Lyapunov exponent (LE), we use the recently developed method based on a parallel evolution of fiducial and auxiliary trajectories, $\psi_f(t)$ and $\psi_a(t)$, under Eq.~(\ref{eq:nonHermHamiltonian}) \cite{yusipov_vershinina_2019}, in the spirit of the classical LE ideology \cite{benettin_galgani_1976}. 
The distance between the trajectories is calculated as the absolute difference between the two corresponding  observables $\theta$. 
We implement a high-performance realization of the quantum jumps method \cite{volokitin_2017} to generate $M_{\mathrm{r}}=10^2$ different trajectories for every considered set of model parameters. 
We first integrate  each trajectory  up to time $t_0 = 10 T$ in order to propagate the model system into the asymptotic regime, and then we follow the dynamics of fiducial and auxiliary trajectories  up to time $t = 10 T$.  

\section{Results}
   
The dynamics of the photonic mode in the periodically modulated in time Kerr-nonlinear cavity \cite{yusipov_vershinina_2020} serves us a reference case and a background to project  our results on. 
Equations (\ref{eq:Hamiltonian})-(\ref{eq:dissipation}) reproduce this case when we set $\delta=0$, $g=0$, $\omega=0$. 
In this case, the spin is decoupled from the photonic subsystem and has no influence on the dynamics of the latter. 
The only relevant dissipative channel is the spontaneous photon emission.
    
\begin{figure}[t]
    \begin{center}
    \includegraphics[width=0.95\columnwidth,keepaspectratio,clip]{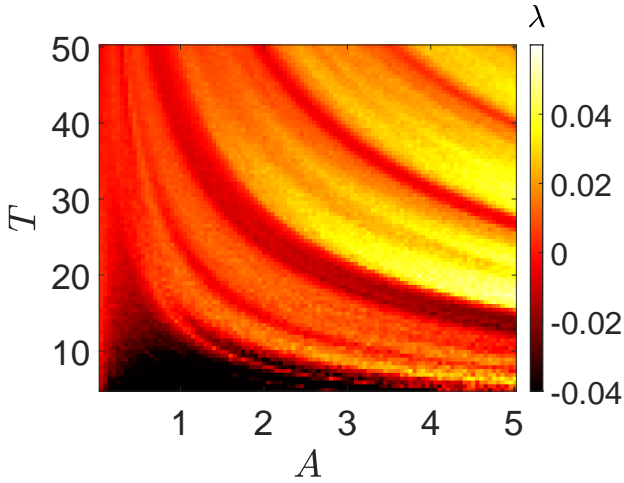}
    \caption{
        (Color online) Largest Lyapunov exponent as a function of the modulation amplitude $A$ and period $T$ in the limit when the spin is absent. 
        The parameters are  $\gamma = 0$, $\delta=0$, $g=0$, and $\omega=0$. 
    }  
    \label{fig:1}
    \end{center}
\end{figure}

Figure \ref{fig:1} shows largest Lyapunov exponent \cite{yusipov_vershinina_2020} $\lambda$ as a function of amplitude $A$ and period $T$. 
For the relatively small values of the modulation period and amplitude, the system remains in the regular regime characterized by negative LE. 
By Increasing each of the control parameters, we can  drive the system into a broad chaotic zone interlaced with narrow tongue-like zones of regular (non-chaotic) dynamics.
    
\begin{figure}[b]
    \begin{center}
    \includegraphics[width=0.95\columnwidth,keepaspectratio,clip]{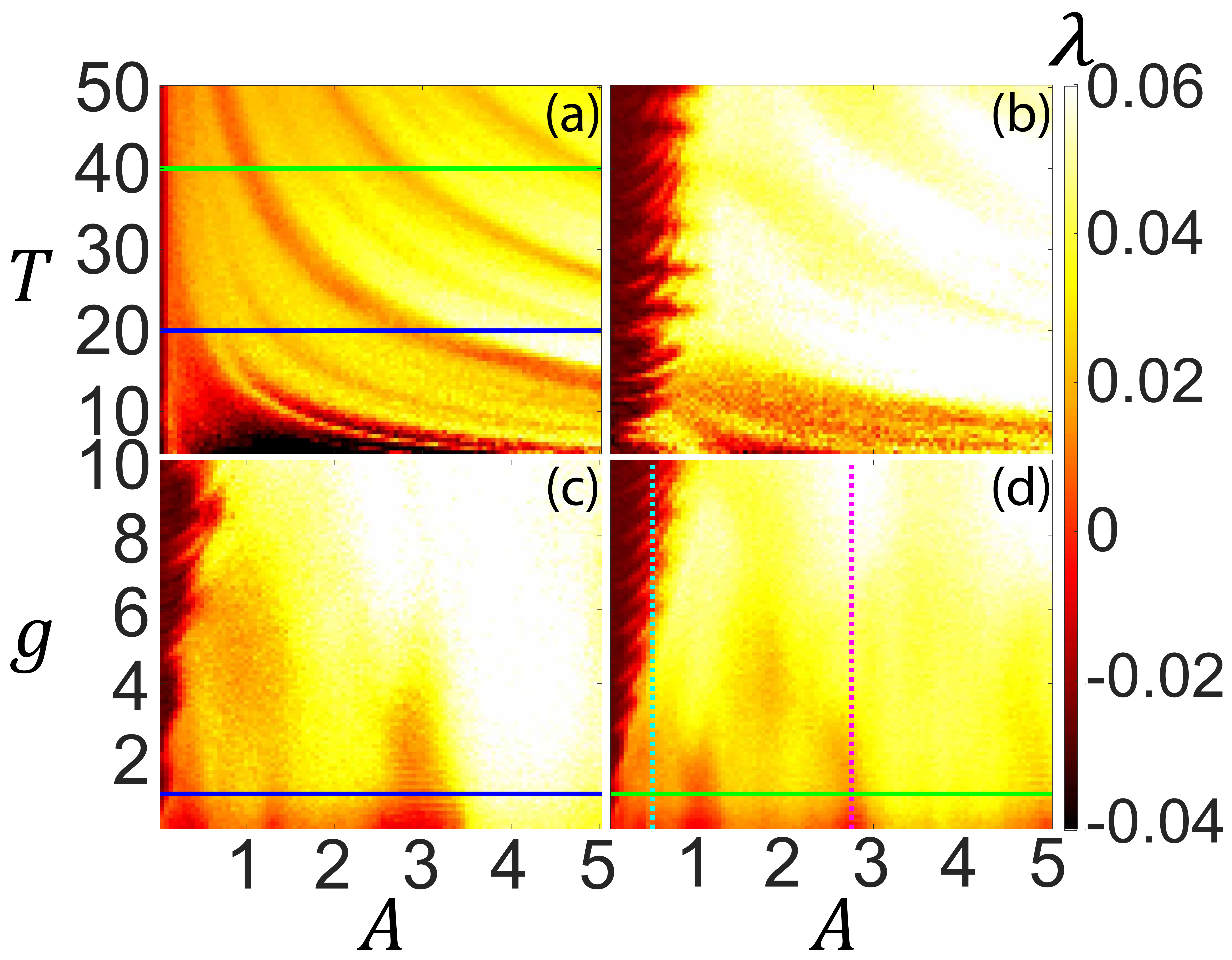}
    \caption{
        Largest Lyapunov exponent as a function of the modulation amplitude $A$ and period $T$ for different value of the spin-photonic coupling $g$.  
        On panels (a-b) the dependencies for $g=1$ (a) and $g=10$ (b) are presented. 
        The blue line on panel (a) corresponds to $T=20$ and the green line to $T=40$. 
        On panels (c-d), the largest Lyapunov exponent is presented as a function of $A$ and $g$ for two different values of the modulation period $T$, $T=20$ (c) and $T=40$ (d). 
        The blue and green lines mark $g=1$, cyan and magenta on (d) correspond to $A=0.5$ and $A=2.75$, respectively (see also Fig.\ref{fig:3}).
    }  
    \label{fig:2}
    \end{center}
\end{figure}

Now we switch on coupling between the spin and photonic subsystems, $g\neq0$, and explore systematically the LE phase diagrams fro different value of teh coupling constant. 
We start from a regime of moderate, $g =1$, spin-photonic interaction. 
Figure \ref{fig:2}(a) shows the largest LE as a function of amplitude $A$ and period $T$  (henceforth we set $\delta=1$ and $\omega=0.05$). 
While the structure of regular and chaotic zones remains essentially intact, the value of the LE increases. 

In other words, by coupling the cavity photonic mode to the spin degree of freedom, we increase the degree of chaos in the dynamics of of the former.
For strong spin-photon coupling, $g=10$, we observe  two new  trends.
Namely, in the strong modulation limit, $A>1$, larger values of $g$ still result in larger LEs.
At the same time, in the case of weak modulations, $A<1$, the increase of the spin-photon coupling leads to 'regularization' of dynamics as it manifested by the negative LE.    

\begin{figure}[t]
    \begin{center}
    \includegraphics[width=0.95\columnwidth,keepaspectratio,clip]{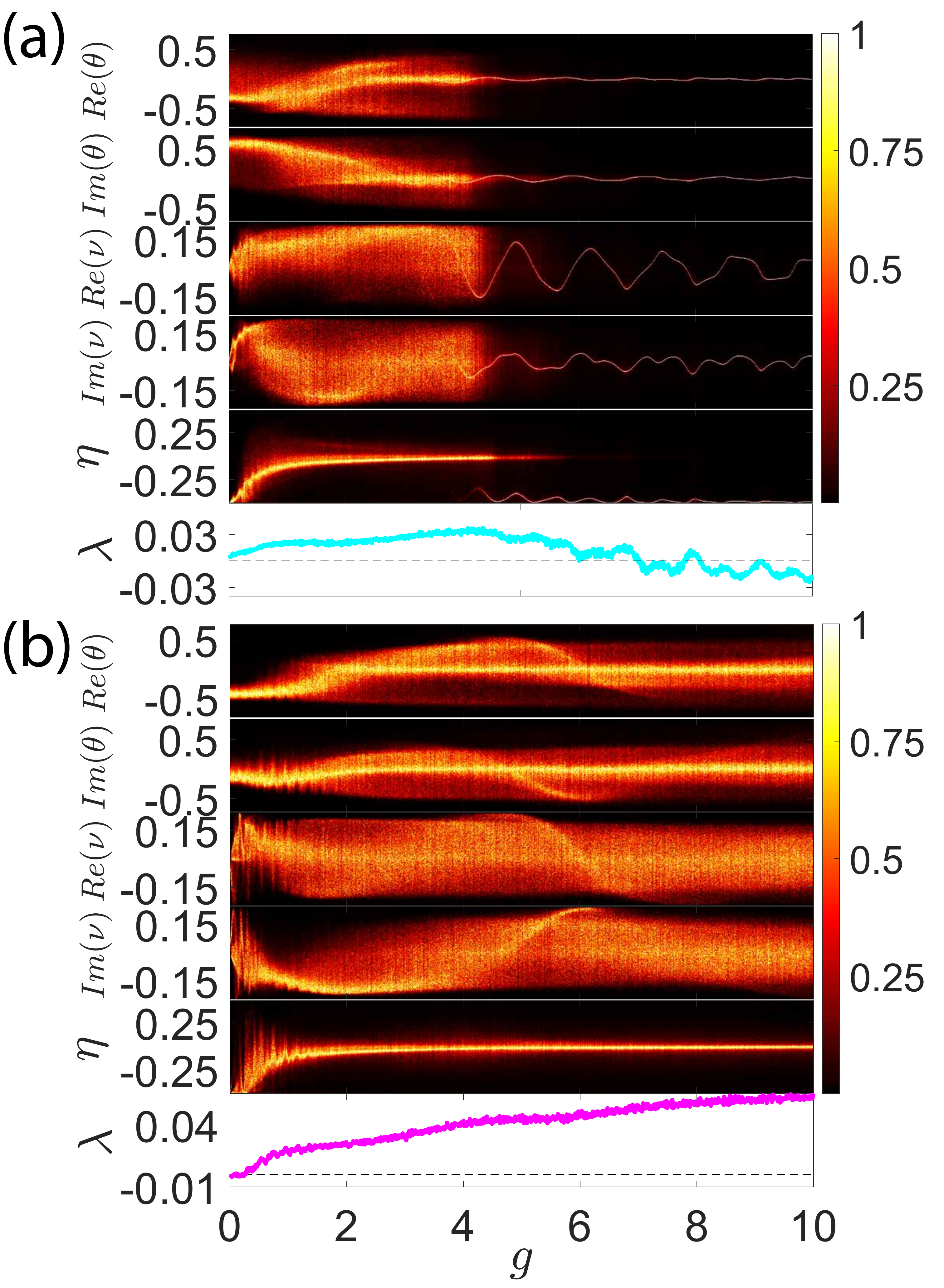}
    \caption{
    	Observables ($\theta, \nu, \eta$) and the largest Lyapunov as a function of spin-photonic coupling $g$ for (a) $A=0.5$ and (b) $A=2.75$ [cyan and magenta lines on Fig.\ref{fig:2}(d)] respectively. 
    	Here $T=40$.
    }  
    \label{fig:3}
    \end{center}
\end{figure}

Next, we fix two values of the modulation period, $T=20$ and $T=40$, and systematically vary the spin-photon coupling $g$.
Figures \ref{fig:2}(b) and \ref{fig:2}(c) show the largest LE as a function of amplitude $A$ and coupling $g$ for $T=20$ and $T=40$ respectively.
The already noted scenarios for weak and strong modulation amplitudes are emerging again.

To get an additional insight, we plot the histograms for observables $(\theta, \nu, \eta)$ for individual quantum trajectories, along with the largest LE versus, fro different values of the  spin-photonic coupling $g$; see Fig.~\ref{fig:3}. 
For the modulation amplitude $A = 0.5$ we observe that the Lyapunov exponent decreases and eventually becomes negative at $g\sim 6\ldots7$.  
Figure \ref{fig:3}(a) illustrates the corresponding change in the dynamics of  the observable, which reduces from the complex evolution to a fixed point.

In contrast, for $A = 2.75$, we observe that the initially regular regime (which starts from $g = 0$), becomes chaotic with the increase of the  coupling strength so that the LE first becomes positive and then gradually increases; see Fig.~\ref{fig:3}(b). 

\begin{figure}[b]
    \begin{center}
    \includegraphics[width=\columnwidth,keepaspectratio,clip]{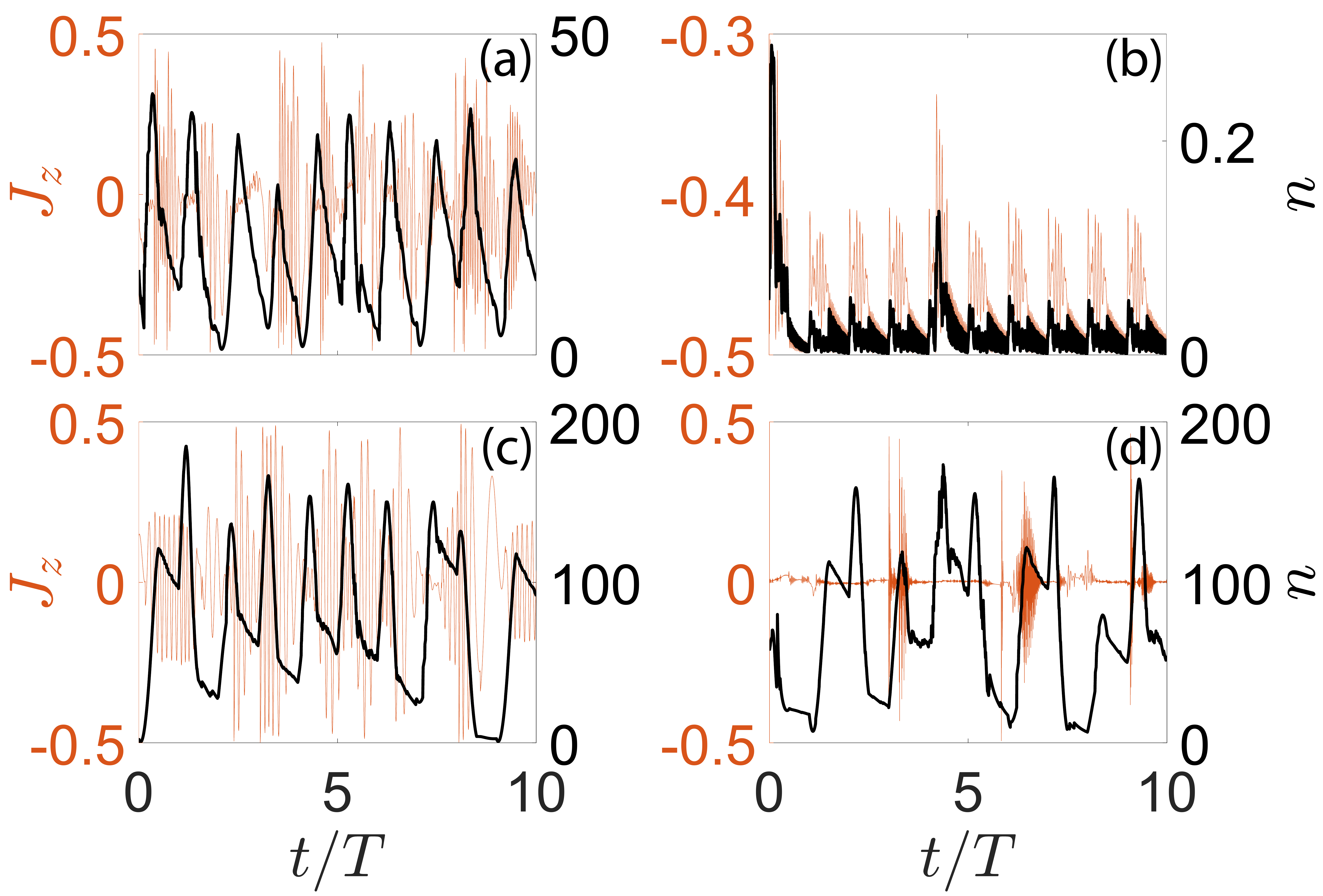}
    \caption{
    	(Color online) Dynamics of spin operator $J_z$ and photon number $n$ expectation values for individual quantum trajectories. 
    	Transition from chaos to regular dynamics due to spin-induced photon leaking for weak modulation $A=0.5$ and different coupling strength, (a) $g=1$ and (b) $g=10$. 
    	Increasing irregularity following the growth of spin-photon coupling from $g=1$ (c) to $g=10$, in the limit of strong modulation, $A=3$. 
    	Here $T=40$. 
    }  
    \label{fig:4}
    \end{center}
\end{figure}

Analysis of a single-trajectory dynamics of the expectation values of spin $J_z$ and photon  $\hat{n}$ operators (see Fig.~\ref{fig:4}) allows to understand the observations. 
For small modulation amplitude, $A=0.5$, it is noteworthy that the transition from a highly chaotic dynamics at $g=1$, to an almost periodic behavior at $g=10$, is characterized by a fast escape of photons from the cavity; see Fig.~\ref{fig:4}(a,b). 
An increase of the expectation number of photons almost immediately leads to an excitation of the spin, which then quickly relaxes back to the groundstate. 
Stronger modulation amplitudes are needed in order to compensate losses through this dissipation channel; see panels (c,d).

The chaotic and regular dynamics of the system has been demonstrated by using the idea of maximal Lyapunov exponents for individual trajectories \cite{yusipov_vershinina_2020}. 
Measurement of these exponents in an experiment is hardly possible. 
Instead we can use the statistics of the delay (waiting time) between the two consecutive photon emission (which is accessible in an experiment \cite{carmichael_1993}) and try to use the corresponding probability density function (pdf) $PDF(\Delta t)$ as an indicator of chaos in the intra-cavity dynamics.

\begin{figure}[t]
    \begin{center}
    \includegraphics[width=0.95\columnwidth,keepaspectratio,clip]{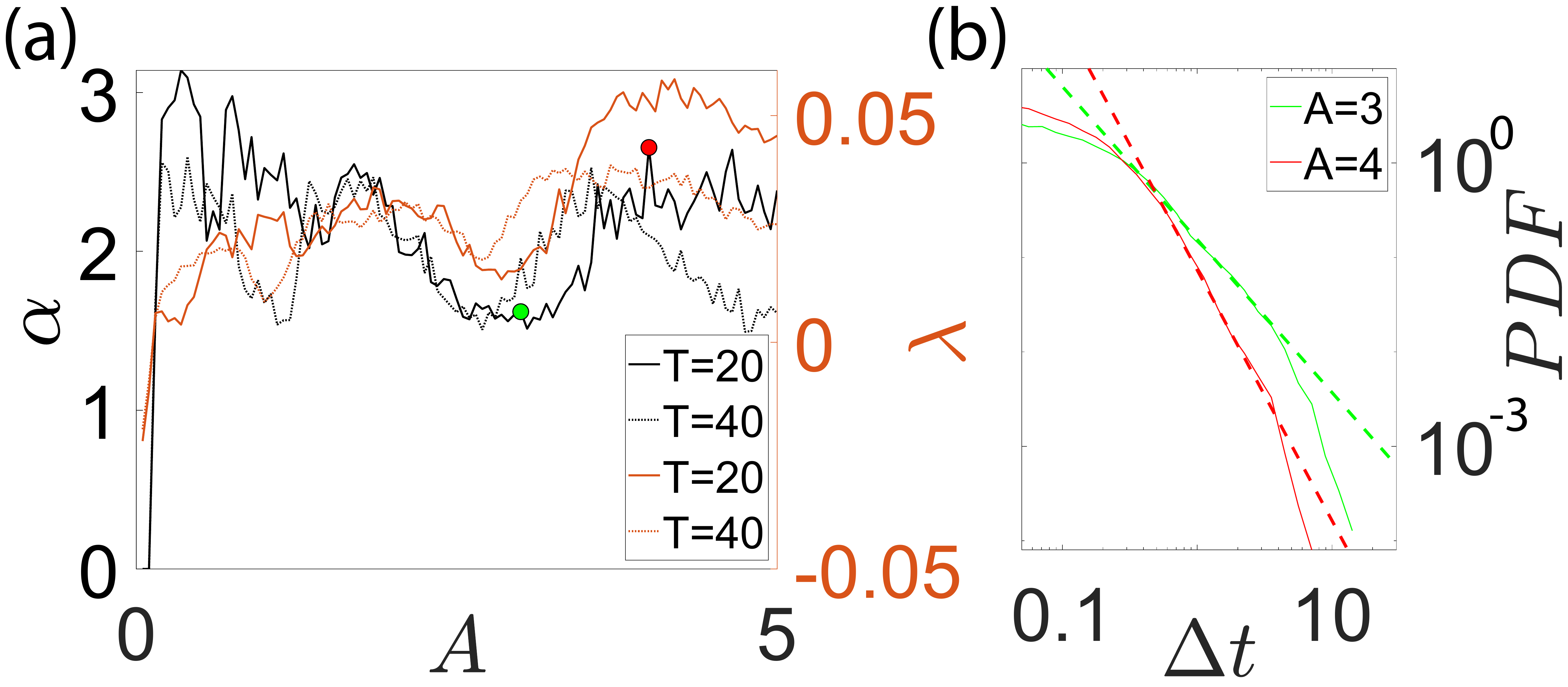}
    \caption{
        (Color online) LE and the power-law fit exponent $\alpha$ for the probability distribution of time intervals between cavity photon emissions (a), $PDF(\Delta t)\sim t^{-\alpha}$ in dependence on modulation amplitude, $A$ (b). 
        Here $g=1$. 
    }  
    \label{fig:5}
    \end{center}
\end{figure}

In our prior work \cite{yusipov_vershinina_2020}, we found that the transitions to chaotic photonic regimes in an open Kerr-nonlinear cavity are associated with appearance of power-law intermediate asymptotics, $PDF(\Delta t)\sim \Delta t^{-\alpha}$, in the corresponding waiting time pdf (which otherwise decays exponentially) \cite{yusipov_vershinina_2020}.
We expect that this effect is generic and will emerge also in the considered spin-photonic model. 
Indeed, the power-law exponent $\alpha$ estimated by using the procedure described in Ref.~\cite{yusipov_vershinina_2020}, is in a nice agreement with the LE chaos quantifier; see Fig.~\ref{fig:5}(b).

\section{Conclusions}
% Cavity QED systems become a popular theoretical and experimental testbed to investigate complex non-equilibrium quantum regimes and their potential in applications, to name quantum computations and metrology.
% Such systems are inherently open and subject to decoherence effects, which, in particular, leads to the novel quantum chaotic phenomena, in the dissipative framework. 
	
We considered an experimentally relevant quantum model, in which complex dynamics appears as a result of the interplay between periodic modulations, different dissipative mechanisms, and interaction between two sub-systems, large (photons) and small (spin) ones. 
Previously, it has been demonstrated that non-equilibrium photonic dynamics, emerging in an open and periodically pumped Kerr-nonlinear cavity, can exhibit dissipative chaotic regimes characterized by positive quantum  Lyapunov exponents \cite{yusipov_vershinina_2020}. 
Here we demonstrated  that a spin-photon interaction  can significantly  modify the intra-cavity dynamics. In the weak modulation limit,  the collective spin-photonic states become less chaotic due to an additional dissipative channel (spin relaxation) which facilitates the photon leakage. 
In contrast, strong modulations can  counterbalance these losses so that  the degree of chaos in the intra-cavity dynamics increases. 
Remarkably that, similar to a simple set-up considered before \cite{yusipov_vershinina_2020}, the degree of chaos can be estimated in an experiment by analyzing  statistics of the photon emission waiting times. 
	
Further generalization to a model with several interacting spins, could provide  a possibility to introduce new factors such as synchronization \cite{xu_tieri_2014}.

\begin{acknowledgments}
The authors acknowledge support of Basis Foundation grant No. 17-12-279-1 and Russian Foundation for Basic Research No. 18-32-20221. 
Numerical simulations were performed on the supercomputers of the Lobachevsky University of Nizhny Novgorod and Moscow State University.
\end{acknowledgments}

\nocite{*}
\bibliography{spin}% Produces the bibliography via BibTeX.

\providecommand{\noopsort}[1]{}\providecommand{\singleletter}[1]{#1}%
\begin{thebibliography}{27}%
\makeatletter
\providecommand \@ifxundefined [1]{%
 \@ifx{#1\undefined}
}%
\providecommand \@ifnum [1]{%
 \ifnum #1\expandafter \@firstoftwo
 \else \expandafter \@secondoftwo
 \fi
}%
\providecommand \@ifx [1]{%
 \ifx #1\expandafter \@firstoftwo
 \else \expandafter \@secondoftwo
 \fi
}%
\providecommand \natexlab [1]{#1}%
\providecommand \enquote  [1]{``#1''}%
\providecommand \bibnamefont  [1]{#1}%
\providecommand \bibfnamefont [1]{#1}%
\providecommand \citenamefont [1]{#1}%
\providecommand \href@noop [0]{\@secondoftwo}%
\providecommand \href [0]{\begingroup \@sanitize@url \@href}%
\providecommand \@href[1]{\@@startlink{#1}\@@href}%
\providecommand \@@href[1]{\endgroup#1\@@endlink}%
\providecommand \@sanitize@url [0]{\catcode `\\12\catcode `\$12\catcode
  `\&12\catcode `\#12\catcode `\^12\catcode `\_12\catcode `\%12\relax}%
\providecommand \@@startlink[1]{}%
\providecommand \@@endlink[0]{}%
\providecommand \url  [0]{\begingroup\@sanitize@url \@url }%
\providecommand \@url [1]{\endgroup\@href {#1}{\urlprefix }}%
\providecommand \urlprefix  [0]{URL }%
\providecommand \Eprint [0]{\href }%
\providecommand \doibase [0]{http://dx.doi.org/}%
\providecommand \selectlanguage [0]{\@gobble}%
\providecommand \bibinfo  [0]{\@secondoftwo}%
\providecommand \bibfield  [0]{\@secondoftwo}%
\providecommand \translation [1]{[#1]}%
\providecommand \BibitemOpen [0]{}%
\providecommand \bibitemStop [0]{}%
\providecommand \bibitemNoStop [0]{.\EOS\space}%
\providecommand \EOS [0]{\spacefactor3000\relax}%
\providecommand \BibitemShut  [1]{\csname bibitem#1\endcsname}%
\let\auto@bib@innerbib\@empty
%</preamble>
\bibitem [{\citenamefont {Ott}(2002)}]{ott_2002}%
  \BibitemOpen
  \bibfield  {author} {\bibinfo {author} {\bibfnamefont {E.}~\bibnamefont
  {Ott}},\ }\href@noop {} {\emph {\bibinfo {title} {Chaos in dynamical
  systems}}}\ (\bibinfo  {publisher} {Cambridge University Press},\ \bibinfo
  {year} {2002})\BibitemShut {NoStop}%
\bibitem [{\citenamefont {Spohn}(1980)}]{spohn_1980}%
  \BibitemOpen
  \bibfield  {author} {\bibinfo {author} {\bibfnamefont {H.}~\bibnamefont
  {Spohn}},\ }\bibfield  {title} {\enquote {\bibinfo {title} {Kinetic equations
  from hamiltonian dynamics: Markovian limits},}\ }\href@noop {} {\bibfield
  {journal} {\bibinfo  {journal} {Reviews of Modern Physics}\ }\textbf
  {\bibinfo {volume} {52}},\ \bibinfo {pages} {569} (\bibinfo {year}
  {1980})}\BibitemShut {NoStop}%
\bibitem [{\citenamefont {Kadanoff}(2007)}]{kadanoff_2007}%
  \BibitemOpen
  \bibfield  {author} {\bibinfo {author} {\bibfnamefont {L.~P.}\ \bibnamefont
  {Kadanoff}},\ }\href@noop {} {\emph {\bibinfo {title} {Statistical physics:
  statics, dynamics and renormalization}}}\ (\bibinfo  {publisher} {World
  Scientific},\ \bibinfo {year} {2007})\BibitemShut {NoStop}%
\bibitem [{\citenamefont {Breuer}\ and\ \citenamefont
  {Petruccione}(2010)}]{breuer_petruccione_2010}%
  \BibitemOpen
  \bibfield  {author} {\bibinfo {author} {\bibfnamefont {H.-P.}\ \bibnamefont
  {Breuer}}\ and\ \bibinfo {author} {\bibfnamefont {F.}~\bibnamefont
  {Petruccione}},\ }\href@noop {} {\emph {\bibinfo {title} {The theory of open
  quantum systems}}}\ (\bibinfo  {publisher} {Oxford Univ. Press},\ \bibinfo
  {year} {2010})\BibitemShut {NoStop}%
\bibitem [{\citenamefont {Altland}\ and\ \citenamefont
  {Haake}(2012)}]{altland_haake_2012}%
  \BibitemOpen
  \bibfield  {author} {\bibinfo {author} {\bibfnamefont {A.}~\bibnamefont
  {Altland}}\ and\ \bibinfo {author} {\bibfnamefont {F.}~\bibnamefont
  {Haake}},\ }\bibfield  {title} {\enquote {\bibinfo {title} {Quantum chaos and
  effective thermalization},}\ }\href@noop {} {\bibfield  {journal} {\bibinfo
  {journal} {Physical Review Letters}\ }\textbf {\bibinfo {volume} {108}},\
  \bibinfo {pages} {073601} (\bibinfo {year} {2012})}\BibitemShut {NoStop}%
\bibitem [{\citenamefont {Chávez-Carlos}\ \emph {et~al.}(2016)\citenamefont
  {Chávez-Carlos}, \citenamefont {Bastarrachea-Magnani}, \citenamefont
  {Lerma-Hernández},\ and\ \citenamefont
  {Hirsch}}]{chavez-carlos_bastarrachea-magnani_2016}%
  \BibitemOpen
  \bibfield  {author} {\bibinfo {author} {\bibfnamefont {J.}~\bibnamefont
  {Chávez-Carlos}}, \bibinfo {author} {\bibfnamefont {M.~A.}\ \bibnamefont
  {Bastarrachea-Magnani}}, \bibinfo {author} {\bibfnamefont {S.}~\bibnamefont
  {Lerma-Hernández}}, \ and\ \bibinfo {author} {\bibfnamefont {J.~G.}\
  \bibnamefont {Hirsch}},\ }\bibfield  {title} {\enquote {\bibinfo {title}
  {Classical chaos in atom-field systems},}\ }\href@noop {} {\bibfield
  {journal} {\bibinfo  {journal} {Physical Review E}\ }\textbf {\bibinfo
  {volume} {94}},\ \bibinfo {pages} {022209} (\bibinfo {year}
  {2016})}\BibitemShut {NoStop}%
\bibitem [{\citenamefont {Chávez-Carlos}\ \emph {et~al.}(2019)\citenamefont
  {Chávez-Carlos}, \citenamefont {López-Del-Carpio}, \citenamefont
  {Bastarrachea-Magnani}, \citenamefont {Stránský}, \citenamefont
  {Lerma-Hernández}, \citenamefont {Santos},\ and\ \citenamefont
  {Hirsch}}]{chavez-carlos_lopez-del-carpio_2019}%
  \BibitemOpen
  \bibfield  {author} {\bibinfo {author} {\bibfnamefont {J.}~\bibnamefont
  {Chávez-Carlos}}, \bibinfo {author} {\bibfnamefont {B.}~\bibnamefont
  {López-Del-Carpio}}, \bibinfo {author} {\bibfnamefont {M.~A.}\ \bibnamefont
  {Bastarrachea-Magnani}}, \bibinfo {author} {\bibfnamefont {P.}~\bibnamefont
  {Stránský}}, \bibinfo {author} {\bibfnamefont {S.}~\bibnamefont
  {Lerma-Hernández}}, \bibinfo {author} {\bibfnamefont {L.~F.}\ \bibnamefont
  {Santos}}, \ and\ \bibinfo {author} {\bibfnamefont {J.~G.}\ \bibnamefont
  {Hirsch}},\ }\bibfield  {title} {\enquote {\bibinfo {title} {Quantum and
  classical lyapunov exponents in atom-field interaction systems},}\
  }\href@noop {} {\bibfield  {journal} {\bibinfo  {journal} {Physical Review
  Letters}\ }\textbf {\bibinfo {volume} {122}},\ \bibinfo {pages} {024101}
  (\bibinfo {year} {2019})}\BibitemShut {NoStop}%
\bibitem [{\citenamefont {Schiró}\ \emph {et~al.}(2016)\citenamefont
  {Schiró}, \citenamefont {Joshi}, \citenamefont {Bordyuh}, \citenamefont
  {Fazio}, \citenamefont {Keeling},\ and\ \citenamefont
  {Türeci}}]{schiro_joshi_bordyuh_2016}%
  \BibitemOpen
  \bibfield  {author} {\bibinfo {author} {\bibfnamefont {M.}~\bibnamefont
  {Schiró}}, \bibinfo {author} {\bibfnamefont {C.}~\bibnamefont {Joshi}},
  \bibinfo {author} {\bibfnamefont {M.}~\bibnamefont {Bordyuh}}, \bibinfo
  {author} {\bibfnamefont {R.}~\bibnamefont {Fazio}}, \bibinfo {author}
  {\bibfnamefont {J.}~\bibnamefont {Keeling}}, \ and\ \bibinfo {author}
  {\bibfnamefont {H.}~\bibnamefont {Türeci}},\ }\bibfield  {title} {\enquote
  {\bibinfo {title} {Exotic attractors of the nonequilibrium rabi-hubbard
  model},}\ }\href@noop {} {\bibfield  {journal} {\bibinfo  {journal} {Physical
  Review Letters}\ }\textbf {\bibinfo {volume} {116}},\ \bibinfo {pages}
  {143603} (\bibinfo {year} {2016})}\BibitemShut {NoStop}%
\bibitem [{\citenamefont {Xu}\ \emph {et~al.}(2014)\citenamefont {Xu},
  \citenamefont {Tieri}, \citenamefont {Fine}, \citenamefont {Thompson},\ and\
  \citenamefont {Holland}}]{xu_tieri_2014}%
  \BibitemOpen
  \bibfield  {author} {\bibinfo {author} {\bibfnamefont {M.}~\bibnamefont
  {Xu}}, \bibinfo {author} {\bibfnamefont {D.}~\bibnamefont {Tieri}}, \bibinfo
  {author} {\bibfnamefont {E.}~\bibnamefont {Fine}}, \bibinfo {author}
  {\bibfnamefont {J.~K.}\ \bibnamefont {Thompson}}, \ and\ \bibinfo {author}
  {\bibfnamefont {M.}~\bibnamefont {Holland}},\ }\bibfield  {title} {\enquote
  {\bibinfo {title} {Synchronization of two ensembles of atoms},}\ }\href@noop
  {} {\bibfield  {journal} {\bibinfo  {journal} {Physical Review Letters}\
  }\textbf {\bibinfo {volume} {113}},\ \bibinfo {pages} {154101} (\bibinfo
  {year} {2014})}\BibitemShut {NoStop}%
\bibitem [{\citenamefont {Hartmann}\ \emph {et~al.}(2017)\citenamefont
  {Hartmann}, \citenamefont {Poletti}, \citenamefont {Ivanchenko},
  \citenamefont {Denisov},\ and\ \citenamefont
  {Hänggi}}]{hartmann_poletti_2017}%
  \BibitemOpen
  \bibfield  {author} {\bibinfo {author} {\bibfnamefont {M.}~\bibnamefont
  {Hartmann}}, \bibinfo {author} {\bibfnamefont {D.}~\bibnamefont {Poletti}},
  \bibinfo {author} {\bibfnamefont {M.}~\bibnamefont {Ivanchenko}}, \bibinfo
  {author} {\bibfnamefont {S.}~\bibnamefont {Denisov}}, \ and\ \bibinfo
  {author} {\bibfnamefont {P.}~\bibnamefont {Hänggi}},\ }\bibfield  {title}
  {\enquote {\bibinfo {title} {Asymptotic floquet states of open quantum
  systems: the role of interaction},}\ }\href@noop {} {\bibfield  {journal}
  {\bibinfo  {journal} {New Journal of Physics}\ }\textbf {\bibinfo {volume}
  {19}},\ \bibinfo {pages} {083011} (\bibinfo {year} {2017})}\BibitemShut
  {NoStop}%
\bibitem [{\citenamefont {Patra}, \citenamefont {Altshuler},\ and\
  \citenamefont {Yuzbashyan}(2019)}]{patra_altshuler_2019}%
  \BibitemOpen
  \bibfield  {author} {\bibinfo {author} {\bibfnamefont {A.}~\bibnamefont
  {Patra}}, \bibinfo {author} {\bibfnamefont {B.~L.}\ \bibnamefont
  {Altshuler}}, \ and\ \bibinfo {author} {\bibfnamefont {E.~A.}\ \bibnamefont
  {Yuzbashyan}},\ }\bibfield  {title} {\enquote {\bibinfo {title} {Chaotic
  synchronization between atomic clocks},}\ }\href@noop {} {\bibfield
  {journal} {\bibinfo  {journal} {Physical Review A}\ }\textbf {\bibinfo
  {volume} {100}},\ \bibinfo {pages} {023418} (\bibinfo {year}
  {2019})}\BibitemShut {NoStop}%
\bibitem [{\citenamefont {Yusipov}\ \emph {et~al.}(2019)\citenamefont
  {Yusipov}, \citenamefont {Vershinina}, \citenamefont {Denisov}, \citenamefont
  {Kuznetsov},\ and\ \citenamefont {Ivanchenko}}]{yusipov_vershinina_2019}%
  \BibitemOpen
  \bibfield  {author} {\bibinfo {author} {\bibfnamefont {I.~I.}\ \bibnamefont
  {Yusipov}}, \bibinfo {author} {\bibfnamefont {O.~S.}\ \bibnamefont
  {Vershinina}}, \bibinfo {author} {\bibfnamefont {S.}~\bibnamefont {Denisov}},
  \bibinfo {author} {\bibfnamefont {S.~P.}\ \bibnamefont {Kuznetsov}}, \ and\
  \bibinfo {author} {\bibfnamefont {M.~V.}\ \bibnamefont {Ivanchenko}},\
  }\bibfield  {title} {\enquote {\bibinfo {title} {Quantum lyapunov exponents
  beyond continuous measurements},}\ }\href@noop {} {\bibfield  {journal}
  {\bibinfo  {journal} {Chaos: An Interdisciplinary Journal of Nonlinear
  Science}\ }\textbf {\bibinfo {volume} {29}},\ \bibinfo {pages} {063130}
  (\bibinfo {year} {2019})}\BibitemShut {NoStop}%
\bibitem [{\citenamefont {Walther}\ \emph {et~al.}(2006)\citenamefont
  {Walther}, \citenamefont {Varcoe}, \citenamefont {Englert},\ and\
  \citenamefont {Becker}}]{walther_varcoe_2006}%
  \BibitemOpen
  \bibfield  {author} {\bibinfo {author} {\bibfnamefont {H.}~\bibnamefont
  {Walther}}, \bibinfo {author} {\bibfnamefont {B.~T.~H.}\ \bibnamefont
  {Varcoe}}, \bibinfo {author} {\bibfnamefont {B.-G.}\ \bibnamefont {Englert}},
  \ and\ \bibinfo {author} {\bibfnamefont {T.}~\bibnamefont {Becker}},\
  }\bibfield  {title} {\enquote {\bibinfo {title} {Cavity quantum
  electrodynamics},}\ }\href@noop {} {\bibfield  {journal} {\bibinfo  {journal}
  {Reports on Progress in Physics}\ }\textbf {\bibinfo {volume} {69}},\
  \bibinfo {pages} {1325} (\bibinfo {year} {2006})}\BibitemShut {NoStop}%
\bibitem [{\citenamefont {Arakawa}\ \emph {et~al.}(2015)\citenamefont
  {Arakawa}, \citenamefont {Finley}, \citenamefont {Gross}, \citenamefont
  {Laussy}, \citenamefont {Solano},\ and\ \citenamefont
  {Vuckovic}}]{arakawa_finley_2015}%
  \BibitemOpen
  \bibfield  {author} {\bibinfo {author} {\bibfnamefont {Y.}~\bibnamefont
  {Arakawa}}, \bibinfo {author} {\bibfnamefont {J.}~\bibnamefont {Finley}},
  \bibinfo {author} {\bibfnamefont {R.}~\bibnamefont {Gross}}, \bibinfo
  {author} {\bibfnamefont {F.}~\bibnamefont {Laussy}}, \bibinfo {author}
  {\bibfnamefont {E.}~\bibnamefont {Solano}}, \ and\ \bibinfo {author}
  {\bibfnamefont {J.}~\bibnamefont {Vuckovic}},\ }\bibfield  {title} {\enquote
  {\bibinfo {title} {Focus on cavity and circuit quantum electrodynamics in
  solids},}\ }\href@noop {} {\bibfield  {journal} {\bibinfo  {journal} {New
  Journal of Physics}\ }\textbf {\bibinfo {volume} {17}},\ \bibinfo {pages}
  {010201} (\bibinfo {year} {2015})}\BibitemShut {NoStop}%
\bibitem [{\citenamefont {Reithmaier}\ \emph {et~al.}(2004)\citenamefont
  {Reithmaier}, \citenamefont {Sęk}, \citenamefont {Löffler}, \citenamefont
  {Hofmann}, \citenamefont {Kuhn}, \citenamefont {Reitzenstein}, \citenamefont
  {Keldysh}, \citenamefont {Kulakovskii}, \citenamefont {Reinecke},\ and\
  \citenamefont {Forchel}}]{reithmaier_sek_2004}%
  \BibitemOpen
  \bibfield  {author} {\bibinfo {author} {\bibfnamefont {J.~P.}\ \bibnamefont
  {Reithmaier}}, \bibinfo {author} {\bibfnamefont {G.}~\bibnamefont {Sęk}},
  \bibinfo {author} {\bibfnamefont {A.}~\bibnamefont {Löffler}}, \bibinfo
  {author} {\bibfnamefont {C.}~\bibnamefont {Hofmann}}, \bibinfo {author}
  {\bibfnamefont {S.}~\bibnamefont {Kuhn}}, \bibinfo {author} {\bibfnamefont
  {S.}~\bibnamefont {Reitzenstein}}, \bibinfo {author} {\bibfnamefont {L.~V.}\
  \bibnamefont {Keldysh}}, \bibinfo {author} {\bibfnamefont {V.~D.}\
  \bibnamefont {Kulakovskii}}, \bibinfo {author} {\bibfnamefont {T.~L.}\
  \bibnamefont {Reinecke}}, \ and\ \bibinfo {author} {\bibfnamefont
  {A.}~\bibnamefont {Forchel}},\ }\bibfield  {title} {\enquote {\bibinfo
  {title} {Strong coupling in a single quantum dot–semiconductor microcavity
  system},}\ }\href@noop {} {\bibfield  {journal} {\bibinfo  {journal}
  {Nature}\ }\textbf {\bibinfo {volume} {432}},\ \bibinfo {pages} {197}
  (\bibinfo {year} {2004})}\BibitemShut {NoStop}%
\bibitem [{\citenamefont {Hennessy}\ \emph {et~al.}(2007)\citenamefont
  {Hennessy}, \citenamefont {Badolato}, \citenamefont {Winger}, \citenamefont
  {Gerace}, \citenamefont {Atatüre}, \citenamefont {Gulde}, \citenamefont
  {Fält}, \citenamefont {Hu},\ and\ \citenamefont
  {Imamoğlu}}]{hennessy_badolato_2007}%
  \BibitemOpen
  \bibfield  {author} {\bibinfo {author} {\bibfnamefont {K.}~\bibnamefont
  {Hennessy}}, \bibinfo {author} {\bibfnamefont {A.}~\bibnamefont {Badolato}},
  \bibinfo {author} {\bibfnamefont {M.}~\bibnamefont {Winger}}, \bibinfo
  {author} {\bibfnamefont {D.}~\bibnamefont {Gerace}}, \bibinfo {author}
  {\bibfnamefont {M.}~\bibnamefont {Atatüre}}, \bibinfo {author}
  {\bibfnamefont {S.}~\bibnamefont {Gulde}}, \bibinfo {author} {\bibfnamefont
  {S.}~\bibnamefont {Fält}}, \bibinfo {author} {\bibfnamefont {E.~L.}\
  \bibnamefont {Hu}}, \ and\ \bibinfo {author} {\bibfnamefont {A.}~\bibnamefont
  {Imamoğlu}},\ }\bibfield  {title} {\enquote {\bibinfo {title} {Quantum
  nature of a strongly coupled single quantum dot–cavity system},}\
  }\href@noop {} {\bibfield  {journal} {\bibinfo  {journal} {Nature}\ }\textbf
  {\bibinfo {volume} {445}},\ \bibinfo {pages} {896} (\bibinfo {year}
  {2007})}\BibitemShut {NoStop}%
\bibitem [{\citenamefont {Yusipov}\ \emph {et~al.}(2020)\citenamefont
  {Yusipov}, \citenamefont {Vershinina}, \citenamefont {Denisov},\ and\
  \citenamefont {Ivanchenko}}]{yusipov_vershinina_2020}%
  \BibitemOpen
  \bibfield  {author} {\bibinfo {author} {\bibfnamefont {I.~I.}\ \bibnamefont
  {Yusipov}}, \bibinfo {author} {\bibfnamefont {O.~S.}\ \bibnamefont
  {Vershinina}}, \bibinfo {author} {\bibfnamefont {S.~V.}\ \bibnamefont
  {Denisov}}, \ and\ \bibinfo {author} {\bibfnamefont {M.~V.}\ \bibnamefont
  {Ivanchenko}},\ }\bibfield  {title} {\enquote {\bibinfo {title} {Photon
  waiting-time distributions: A keyhole into dissipative quantum chaos},}\
  }\href@noop {} {\bibfield  {journal} {\bibinfo  {journal} {Chaos: An
  Interdisciplinary Journal of Nonlinear Science}\ }\textbf {\bibinfo {volume}
  {30}},\ \bibinfo {pages} {023107} (\bibinfo {year} {2020})}\BibitemShut
  {NoStop}%
\bibitem [{\citenamefont {Spiller}\ and\ \citenamefont
  {Ralph}(1994)}]{spiller_ralph_1994}%
  \BibitemOpen
  \bibfield  {author} {\bibinfo {author} {\bibfnamefont {T.}~\bibnamefont
  {Spiller}}\ and\ \bibinfo {author} {\bibfnamefont {J.}~\bibnamefont
  {Ralph}},\ }\bibfield  {title} {\enquote {\bibinfo {title} {The emergence of
  chaos in an open quantum system},}\ }\href@noop {} {\bibfield  {journal}
  {\bibinfo  {journal} {Physics Letters A}\ }\textbf {\bibinfo {volume}
  {194}},\ \bibinfo {pages} {235} (\bibinfo {year} {1994})}\BibitemShut
  {NoStop}%
\bibitem [{\citenamefont {Brun}, \citenamefont {Percival},\ and\ \citenamefont
  {Schack}(1996)}]{brun_percival_1996}%
  \BibitemOpen
  \bibfield  {author} {\bibinfo {author} {\bibfnamefont {T.~A.}\ \bibnamefont
  {Brun}}, \bibinfo {author} {\bibfnamefont {I.~C.}\ \bibnamefont {Percival}},
  \ and\ \bibinfo {author} {\bibfnamefont {R.}~\bibnamefont {Schack}},\
  }\bibfield  {title} {\enquote {\bibinfo {title} {Quantum chaos in open
  systems: a quantum state diffusion analysis},}\ }\href@noop {} {\bibfield
  {journal} {\bibinfo  {journal} {Journal of Physics A: Mathematical and
  General}\ }\textbf {\bibinfo {volume} {29}},\ \bibinfo {pages} {2077}
  (\bibinfo {year} {1996})}\BibitemShut {NoStop}%
\bibitem [{\citenamefont {Alicki}\ and\ \citenamefont
  {Lendi}(2007)}]{alicki_lendi_2007}%
  \BibitemOpen
  \bibfield  {author} {\bibinfo {author} {\bibfnamefont {R.}~\bibnamefont
  {Alicki}}\ and\ \bibinfo {author} {\bibfnamefont {K.}~\bibnamefont {Lendi}},\
  }\href@noop {} {\emph {\bibinfo {title} {Quantum Dynamical Semigroups and
  Applications}}}\ (\bibinfo  {publisher} {Springer Berlin Heidelberg},\
  \bibinfo {year} {2007})\BibitemShut {NoStop}%
\bibitem [{\citenamefont {Dum}\ \emph {et~al.}(1992)\citenamefont {Dum},
  \citenamefont {Parkins}, \citenamefont {Zoller},\ and\ \citenamefont
  {Gardiner}}]{dum_parkins_1992}%
  \BibitemOpen
  \bibfield  {author} {\bibinfo {author} {\bibfnamefont {R.}~\bibnamefont
  {Dum}}, \bibinfo {author} {\bibfnamefont {A.~S.}\ \bibnamefont {Parkins}},
  \bibinfo {author} {\bibfnamefont {P.}~\bibnamefont {Zoller}}, \ and\ \bibinfo
  {author} {\bibfnamefont {C.~W.}\ \bibnamefont {Gardiner}},\ }\bibfield
  {title} {\enquote {\bibinfo {title} {Monte carlo simulation of master
  equations in quantum optics for vacuum, thermal, and squeezed reservoirs},}\
  }\href@noop {} {\bibfield  {journal} {\bibinfo  {journal} {Physical Review
  A}\ }\textbf {\bibinfo {volume} {46}},\ \bibinfo {pages} {4382} (\bibinfo
  {year} {1992})}\BibitemShut {NoStop}%
\bibitem [{\citenamefont {Mølmer}, \citenamefont {Castin},\ and\ \citenamefont
  {Dalibard}(1993)}]{molmer_castin_1993}%
  \BibitemOpen
  \bibfield  {author} {\bibinfo {author} {\bibfnamefont {K.}~\bibnamefont
  {Mølmer}}, \bibinfo {author} {\bibfnamefont {Y.}~\bibnamefont {Castin}}, \
  and\ \bibinfo {author} {\bibfnamefont {J.}~\bibnamefont {Dalibard}},\
  }\bibfield  {title} {\enquote {\bibinfo {title} {Monte carlo wave-function
  method in quantum optics},}\ }\href@noop {} {\bibfield  {journal} {\bibinfo
  {journal} {Journal of the Optical Society of America B}\ }\textbf {\bibinfo
  {volume} {10}},\ \bibinfo {pages} {524} (\bibinfo {year} {1993})}\BibitemShut
  {NoStop}%
\bibitem [{\citenamefont {Plenio}\ and\ \citenamefont
  {Knight}(1998)}]{plenio_knight_1998}%
  \BibitemOpen
  \bibfield  {author} {\bibinfo {author} {\bibfnamefont {M.}~\bibnamefont
  {Plenio}}\ and\ \bibinfo {author} {\bibfnamefont {P.}~\bibnamefont
  {Knight}},\ }\bibfield  {title} {\enquote {\bibinfo {title} {The quantum-jump
  approach to dissipative dynamics in quantum optics},}\ }\href@noop {}
  {\bibfield  {journal} {\bibinfo  {journal} {Reviews of Modern Physics}\
  }\textbf {\bibinfo {volume} {70}},\ \bibinfo {pages} {101} (\bibinfo {year}
  {1998})}\BibitemShut {NoStop}%
\bibitem [{\citenamefont {Daley}(2014)}]{daley_2014}%
  \BibitemOpen
  \bibfield  {author} {\bibinfo {author} {\bibfnamefont {A.~J.}\ \bibnamefont
  {Daley}},\ }\bibfield  {title} {\enquote {\bibinfo {title} {Quantum
  trajectories and open many-body quantum systems},}\ }\href@noop {} {\bibfield
   {journal} {\bibinfo  {journal} {Advances in Physics}\ }\textbf {\bibinfo
  {volume} {63}},\ \bibinfo {pages} {77} (\bibinfo {year} {2014})}\BibitemShut
  {NoStop}%
\bibitem [{\citenamefont {Carmichael}(1993)}]{carmichael_1993}%
  \BibitemOpen
  \bibfield  {author} {\bibinfo {author} {\bibfnamefont {H.}~\bibnamefont
  {Carmichael}},\ }\bibfield  {title} {\enquote {\bibinfo {title} {An open
  systems approach to quantum optics},}\ }\href@noop {} {\bibfield  {journal}
  {\bibinfo  {journal} {Lecture Notes in Physics Monographs}\ } (\bibinfo
  {year} {1993})}\BibitemShut {NoStop}%
\bibitem [{\citenamefont {Benettin}, \citenamefont {Galgani},\ and\
  \citenamefont {Strelcyn}(1976)}]{benettin_galgani_1976}%
  \BibitemOpen
  \bibfield  {author} {\bibinfo {author} {\bibfnamefont {G.}~\bibnamefont
  {Benettin}}, \bibinfo {author} {\bibfnamefont {L.}~\bibnamefont {Galgani}}, \
  and\ \bibinfo {author} {\bibfnamefont {J.-M.}\ \bibnamefont {Strelcyn}},\
  }\bibfield  {title} {\enquote {\bibinfo {title} {Kolmogorov entropy and
  numerical experiments},}\ }\href@noop {} {\bibfield  {journal} {\bibinfo
  {journal} {Physical Review A}\ }\textbf {\bibinfo {volume} {14}},\ \bibinfo
  {pages} {2338} (\bibinfo {year} {1976})}\BibitemShut {NoStop}%
\bibitem [{\citenamefont {Volokitin}\ \emph {et~al.}(2017)\citenamefont
  {Volokitin}, \citenamefont {Liniov}, \citenamefont {Meyerov}, \citenamefont
  {Hartmann}, \citenamefont {Ivanchenko}, \citenamefont {Hänggi},\ and\
  \citenamefont {Denisov}}]{volokitin_2017}%
  \BibitemOpen
  \bibfield  {author} {\bibinfo {author} {\bibfnamefont {V.}~\bibnamefont
  {Volokitin}}, \bibinfo {author} {\bibfnamefont {A.}~\bibnamefont {Liniov}},
  \bibinfo {author} {\bibfnamefont {I.}~\bibnamefont {Meyerov}}, \bibinfo
  {author} {\bibfnamefont {M.}~\bibnamefont {Hartmann}}, \bibinfo {author}
  {\bibfnamefont {M.}~\bibnamefont {Ivanchenko}}, \bibinfo {author}
  {\bibfnamefont {P.}~\bibnamefont {Hänggi}}, \ and\ \bibinfo {author}
  {\bibfnamefont {S.}~\bibnamefont {Denisov}},\ }\bibfield  {title} {\enquote
  {\bibinfo {title} {Computation of the asymptotic states of modulated open
  quantum systems with a numerically exact realization of the quantum
  trajectory method},}\ }\href@noop {} {\bibfield  {journal} {\bibinfo
  {journal} {Physical Review E}\ }\textbf {\bibinfo {volume} {96}},\ \bibinfo
  {pages} {053313} (\bibinfo {year} {2017})}\BibitemShut {NoStop}%
\end{thebibliography}%

\end{document}